\documentclass[reqno]{amsart}
\usepackage{amsfonts, amsmath, amssymb, amsthm, amscd, setspace, subfigure}
\usepackage{graphicx}
\usepackage[all,cmtip]{xy}
\usepackage{xr-hyper}
\usepackage[unicode]{hyperref}

\newcommand{\rb}{r _{ 1 } }
\newcommand{\ran}{x _+  }
\newcommand{\rbn}{y_+  }
\newcommand{\rab}{x _-  }
\newcommand{\rbb}{y_-  }

\newcommand{\btnone}{\beta _1 ^+  }
\newcommand{\btntwo}{\beta _2 ^+  }
\newcommand{\btnthree}{\beta _3 ^+}
\newcommand{\bbone}{\beta _1 ^+}
\newcommand{\bbtwo}{\beta _2 ^+  }
\newcommand{\bbthreep}{\beta _3 ^+ }
\newcommand{\bbthreem}{\beta _3 ^-  }

\newcommand{\varphitn}{ \varphi _{ \mathrm{n}}  }
\newcommand{\varphitb}{ \varphi _{ \mathrm{b}}  }
\newcommand{\gtn}{g _{ +  }}
\newcommand{\gtb}{g _{ - }}
\newcommand{\Deltatn}{\Delta _{ + }}
\newcommand{\Deltabp}{\Delta _{-p}}
\newcommand{\Msd}{M_{ +  }}
\newcommand{\Masd}{M_{ - }}
\newcommand{\Deltab}{\Delta _{ - }}
\newcommand{\Deltapm}{\Delta _{ \pm }}

\newcommand{\be}{\beta _3 ^{ \mathrm{exact}}}
\renewcommand{\ae}{\alpha  _3 ^{ \mathrm{exact}}}

\title{Harmonic forms on asymptotically AdS metrics}
\author[Franchetti]{Guido Franchetti}
\author[S\'anchez Gal\'an]{Ra\'ul S\'anchez Gal\'an}
\address[GF]{Department of Mathematical Sciences, University of Bath,
  Claverton Down, Bath BA2 7AY, England, United Kingdom}
\email{\href{mailto: gf424@bath.ac.uk}{gf424[at]bath.ac.uk}, ORCID: \href{https://orcid.org/0000-0002-1511-6204}{0000-0002-1511-6204}}
\address[RSG]{School of Mathematics and Statistics, Nanjing University of Science and
Technology, Xuanwu, Nanjing, Jiangsu 210094, People's Republic of China}

\begin{document}
\maketitle
\begin{abstract}
In this paper we study the rotationally invariant harmonic cohomology of a 2-parameter family of Einstein metrics $g$  which admits a cohomogeneity one action of $SU (2) \times U (1) $ and has AdS asymptotics.   Depending on the values of the parameters, $g$ is either of NUT type, if the fixed-point locus of the $U (1) $ action is 0-dimensional,
or of bolt type, if it is 2-dimensional. We find that if $g$ is of NUT type then the space of $SU (2) $-invariant harmonic 2-forms is 3-dimensional and consists entirely of self-dual forms; if $g$ is of bolt type it is 4-dimensional. In both cases we explicitly determine a basis. The pair  $(g,F)$ for $F$ a self-dual harmonic 2-form is also a solution of the bosonic sector of $4D   $ supergravity. We determine for which choices it is a supersymmetric solution  and the amount of preserved supersymmetry.
\end{abstract}

\section{Introduction}
In this paper we study the harmonic cohomology, that is the space of harmonic $L ^2$  forms, of a certain 2-parameter family of Riemannian Einstein manifolds with negative cosmological constant $(\mathcal{M} ,g )$ of bi-axial Bianchi IX type. We also determine when the pair $(g, F)$, for $F$ an (anti) self-dual harmonic form, gives a supersymmetric solution of the supergravity equations and determine the amount of preserved supersymmetry.

A harmonic 2-form is a solution of the Maxwell equations of electromagnetism, and solutions of the coupled Einstein-Maxwell system are of obvious mathematical and physical interest. In four dimensions,  the reason for physical interest is twofold: not just as solutions of the general relativity equations in the presence of an electromagnetic field, but also as solutions of the formally equivalent equations for  the bosonic sector of $4D $ supergravity, where the Maxwell field describes the graviphoton.

Solutions with negative cosmological constant  are  particularly important because of their relevance to the anti de Sitter/conformal field theory (AdS/CFT) correspondence \cite{Maldacena:1997}. Initially formulated in the conformally flat case, the AdS/CFT correspondence is believed to extend to curved solutions which are  asymptotically locally AdS, that is, which asymptotically approach anti de Sitter  space (or a finite quotient of it) --- in Riemannian signature the name asymptotically locally hyperbolic would  be more appropriate.  The possibility of using supersymmetric localisation techniques  \cite{Pestun2012LocalizationOG,Kapustin:2009kz} to perform  non-perturbative computations on curved Riemannian manifolds  also  motivates the study of curved solutions of the supergravity equations in their role of supersymmetric gravity duals of rigid supersymmetric field theories on curved boundaries \cite{martelli:2012,martelli:2013a,martelli:2013,farquet:2016,Huang:2014gca,Nishioka:2014mwa}.

Reference \cite{martelli:2013} gives an exhaustive treatment of the gravity dual of rigid supersymmetric field theories defined on a bi-axially squashed 3-sphere, that is, a topological 3-sphere with a metric admitting an $SU(2) \times U(1)$ isometric action. It would be interesting to understand the gravity dual in the case of boundaries having $SU(2)$, rather than $SU(2)\times U(1)$, symmetry. While dropping the extra axial symmetry may seem a minor modification, it can have a profound effect on the type of solutions allowed. For example,  two famous  Bianchi IX type hyperk\"ahler metrics are the Taub-NUT and the Atiyah-Hitchin manifolds. While the former has symmetry $SU(2)\times U(1)$, the latter  only arises if the symmetry is relaxed to $SU(2)/ \mathbb{Z}  _2 $.

However finding the relevant supergravity backgrounds, that is global solutions of the Einstein-Maxwell system with $SU(2)$ symmetry and locally AdS asymptotics, already requires a substantial effort, before even being able to consider the amount of supersymmetry these backgrounds have. Therefore, one of the aims of this paper is to test the ground by considering the case where the metric retains the extra $U(1)$ symmetry, but the graviphoton field does not. The results are encouraging: we find harmonic forms which are $SU(2)$- but not $SU(2) \times U(1)$-invariant and which depend in an interesting way on the detailed structure of the metric. We do not obtain new Killing spinors, however it is clear from our computations that this is a consequence of the mismatch in symmetry between the metric and the graviphoton. Once this mismatch is lifted we expect extra solutions, which have no analogue in the axially symmetric case, to arise.

From the mathematical point of view, the study of harmonic objects on a Riemannian manifold is a natural problem, and determining the harmonic cohomology of particular examples  gives useful insight in the case of non-compact manifolds, where the usual Hodge decomposition results do not apply. In fact, not much is known about generic harmonic forms on a non-compact manifold, but the situation improves in the case of square-integrable ones \cite{carron:2007}.  For certain asymptotic behaviours of the metric $g$, the harmonic cohomology of $\mathcal{M} $ can be characterised in terms of a particular compactification \cite{hausel:2004} of $\mathcal{M} $. Important  examples include  ALF gravitational instantons, which have finite-dimensional harmonic cohomology in middle dimension \cite{etesi:2008,franchetti:2014,franchetti2019harmonic,baird:2020}. In contrast,  asymptotically (locally) hyperbolic metrics, such as the one we are going to consider in this paper, are  conformally compact and have infinite dimensional middle dimension harmonic cohomology \cite{mazzeo:1988}. However, we will obtain a finite-dimensional space by restricting to forms invariant under a subgroup of the isometry group.

The outline of this paper is as follows.  In Section \ref{secmetric} we introduce the class of metric that we are going to study: a family $g$ of asymptotically locally hyperbolic Einstein metric of bi-axial Bianchi IX type, that is, admitting a cohomogeneity one action of $SU (2) \times U (1) $. In terms of  the basis $(\eta _i )$ of left-invariant 1-forms on $SU (2) $, $g$ has the form
\begin{equation} 
\begin{split} 
\label{gintro} 
g &= \left( \frac{r ^2 - N ^2 }{\Delta (r) } \right)  \mathrm{d} r^2 + (r ^2 - N ^2 ) (\eta  _1 ^2 + \eta _2 ^2 ) + 4 N ^2 \left( \frac{\Delta (r) }{r ^2 - N ^2 } \right) \eta_3^2,\\
\Delta (r)   &=  r ^4+ (1 -6 N ^2 )r ^2  - 2 M r + N ^2 (1 -3 N ^2 ).
\end{split} 
\end{equation}
In (\ref{gintro}) we have fixed the overall length scale by setting the cosmological constant to $-3 $. The two continuos parameters $M$, $N$  are thus effective. An additional discrete parameter $p>0$ arises by considering $\mathbb{Z}  _p $ quotients of the $U (1) $ orbits.

The metric (\ref{gintro}) is smooth only for particular values of the parameters and we review when this is the case. Two classes of solutions arise, depending on the nature of the fixed point locus of the $U (1) $ action. If it is a point, known as a NUT, one needs to take $p =1 $ and obtains a 1-parameter family of topologically trivial solutions known as Taub-NUT-AdS. If it is a 2-sphere, known as a bolt, then for every value of $p  $ one obtains a 1-parameter family of solutions with the topology of the  complex line bundle $\mathcal{O} (-p )\rightarrow \mathbb{C}  P ^1 $.
 For future use, we also determine when $g$ has an (anti) self-dual Weyl tensor. It turns out that the NUT-type metrics are always self-dual, while for bolt-type metrics we obtain further constraints.

In Section \ref{harmonicforms} we study the  harmonic cohomology of  the spaces introduced in Section \ref{secmetric}. It is non-trivial only in middle dimension, where it is infinite dimensional. We reduce to a finite dimensional subcase by considering rotationally invariant, i.e.~$SU (2) $-invariant, 2-forms. We show, by exhibiting an explicit basis, that the space of $L ^2  $, $SU (2) $-invariant harmonic 2-forms is  3-dimensional and consists entirely of self-dual forms for metrics of NUT type;  4-dimensional with a 3-dimensional self-dual subspace in the case of metrics of bolt-type.
Because of the assumed $SU (2) $ symmetry, finding the harmonic forms reduces to solving an ODE in the radial variable. Forms which are also invariant under the extra $U (1)$ isometry  were already known \cite{martelli:2013}. They are given by  $\mathrm{d} \xi^\flat $ and $* \mathrm{d} \xi ^\flat $  for $\xi$ the Killing vector field generating the extra $U (1) $ isometry and have a  particularly simple expression which does not depend on the polynomial $\Delta (r) $ appearing in (\ref{gintro}). This is not the case for the  non-$U (1) $-invariant forms which depend explicitly on the roots of $\Delta(r) $.
The results in this paper extend those in \cite{franchetti2019harmonic} where the case of vanishing cosmological constant is considered.

In Section \ref{kgspsss} we consider the problem of when the pair $(g, F )$, for $g$ one of the metrics of Section \ref{secmetric} and $F$ a harmonic (anti) self-dual 2-form of the type studied in Section \ref{harmonicforms}, is supersymmetric, that is, admits non-trivial solutions of a certain first order linear PDE known as the Killing spinor equation. Having reviewed some global issues due to the fact that not all the manifolds we consider are spin, and discussed the resulting quantisation conditions on $F$, we proceed to examine the integrability conditions for the Killing spinor equation. 

The integrability conditions  reproduce  the constraint forcing $g$ to be half-conformally-flat  (i.e.~the associated Weyl tensor is either self-dual or anti self-dual), something which was known a priori from the general arguments in \cite{dunajski:2010a}. 
We then proceed to explicitly solve the Killing spinor equations obtaining, for both NUT and bolt type metrics, a  1/4 BPS solution and a 1/2 BPS one. These supersymmetric solutions only arise if the harmonic form $F$  is $U (1) $-invariant and satisfies a certain normalisation condition which, in the topologically non-trivial case of bolt-type metrics, guarantees that $F$ is the curvature of a $U (1) $ connection.

It is important to point out that the problem studied in Section \ref{kgspsss} has been  solved in \cite{martelli:2013} for solutions of the Einstein-Maxwell system having $SU (2) \times U (1) $ invariance and hyperbolic asymptotics. Because of the required $SU (2) \times U (1) $ invariance, the 2-form $F$  considered in \cite{martelli:2013} is less general than the one we consider, see the main text for more details. However, since we find that  non-trivial  Killing spinors only arise  if $F$ is $U (1) $-invariant, thus reducing $F$ to the form considered in \cite{martelli:2013},  we end up re-obtaining their solutions.

\section{Bi-axial Bianchi IX type metrics}
\label{secmetric} 
\subsection{Conventions}

We let $\{\eta_1,\eta_2,\eta_3\}$ be a basis for the space of left-invariant differential 1-forms on $SU(2)$. In terms of the Euler angles  $(\theta , \phi , \psi )$  parametrising $SU (2) $ we have
\begin{equation} \begin{split}  
\label{etas}
\eta_1 &= \sin \psi \, \mathrm{d} \theta - \cos \psi \sin \theta \,\mathrm{d}  \phi, \\
\eta_2 &=  \cos \psi \, \mathrm{d} \theta + \sin \psi \sin \theta \, \mathrm{d}  \phi , \\
\eta_3 &= \mathrm{d}  \psi + \cos \theta \, \mathrm{d}  \phi.
\end{split}
\end{equation} 
We will consider Riemannian orientable manifolds of dimension 4 endowed with a bi-axial Bianchi IX metric. A Bianchi IX metric admits an isometric action of  $SU (2)$ (or its quotient by a finite subgroup) of cohomogeneity one, that is, with  generic orbits of codimension one.  Such a metric can be  locally written  \cite{gibbons1979positive}
\begin{equation} \label{Bianchi_metric}
g= f^2 dr^2 + a^2 \eta_1^2 + b^2 \eta_2^2 + c^2 \eta_3^2,
\end{equation}
where  $f, a, b, c$ are functions of a radial variable $r$ only. 

For a bi-axial Bianchi IX metric  $a =b $ in (\ref{Bianchi_metric}), 
so that
\begin{equation}
g=  f ^2 \, d r ^2 + a ^2 (\eta _1  ^2 + \eta _2  ^2 )+ c ^2 \eta _3 ^2
\end{equation}
and there is an additional  $U (1) $ isometric action generated by the vector field $\partial _\psi $ dual to $\eta _3 $. Note that
\begin{equation}
\mathrm{d}  \Omega ^2 = \eta_1^2 + \eta_2^2 = \mathrm{d}  \theta^2 + \sin^2 \theta \, \mathrm{d}  \phi^2 
\end{equation}
is the round metric on $S ^2 $. 
Fixed point sets of the $U (1) $ action are either points, which are called NUTs, or 2-surfaces, which are called bolts \cite{gibbons:1979b}. Away from fixed points of the $U (1) $ action a hypersurface of fixed $r$ has  the topology of a circle fibration over $S ^2 $.  The usual spherical coordinates $ \theta \in[0, \pi ] $, $\phi \in [0, 2 \pi   )$ parametrise the base $S^2$ and the $S ^1 $ fibre is parametrised by $\psi$. We take $\psi$ to have period
\begin{equation}
T = \frac{4 \pi }{p} 
\end{equation} 
 with $p$ a positive integer. For $p =1 $ the circle fibration is the standard Hopf fibration $S ^3 \rightarrow S ^2 $. For $p > 1 $ we have a lens space $S ^1 \hookrightarrow  S ^3 / \mathbb{Z}  _p \rightarrow S ^2 $ with $\mathbb{Z}  _p $ acting on the $S ^1 $ fibre.

We are going to consider the following family of bi-axial Bianchi IX metrics,
\begin{equation} 
\label{bb9} 
\begin{split}
g&=  f ^2 \, d r ^2 + a ^2 (\eta _1  ^2 + \eta _2  ^2 )+ c ^2 \eta _3 ^2,\\
a &= \sqrt{ r ^2 - N ^2 }, \quad f = - \sqrt{   \frac{r ^2 - N ^2 }{ \Delta (r)} } , \quad c = 2 N \sqrt{ \frac{ \Delta (r) }{r ^2 - N ^2 } },
\end{split}
\end{equation} 
where $\Delta (r) $ is the 4-th order polynomial  
\begin{equation}
\label{deltapol} 
\begin{split} 
\Delta (r)   &= -\frac{ \Lambda }{3 } r ^4+ (1 + 2 \Lambda N ^2 )r ^2  - 2 M r + N ^2 (1 + \Lambda N ^2 ) .
\end{split} 
\end{equation} 
On occasion, we will work with the orthonormal coframe 
\begin{equation} \label{on_cofr_axial}
e^1 = a \eta_1,\quad  e^2 = a \eta_2, \quad e^3 = c \eta_3, \quad e^4 = -f dr. 
\end{equation}
The corresponding volume element is,
\begin{equation} 
\label{dvol_form}
\operatorname{vol }  = e ^1 \wedge e ^2 \wedge e ^3 \wedge e ^4 
=  -2 N a^2 dr \wedge \eta_1 \wedge \eta_2 \wedge \eta_3 = -2 N a^2 \sin \theta \,  dr \wedge d\theta \wedge d \phi \wedge d \psi.  
\end{equation}

The metric (\ref{bb9}) arises as a special case of the construction  in \cite{page:1987}. It is Einstein with cosmological constant $\Lambda$.    The three parameters appearing in (\ref{bb9}), (\ref{deltapol})  are:
\begin{itemize}
\item the NUT parameter $N$,
\item the Komar mass $M$, 
\item the cosmological constant $\Lambda $, which sets a length scale $l$ via
\begin{equation}
l = \sqrt{ \frac{3}{|\Lambda|}}.
\end{equation} 
\end{itemize}

For $\Lambda =0 $ one obtains the Ricci-flat metrics studied in \cite{franchetti2019harmonic}. For  generic values of $M$ and $N$,  these metrics have conical singularities while special values  give the  self-dual Taub-NUT metric, Taub-bolt, Euclidean Schwarzschild and Eguchi-Hanson.
For $\Lambda > 0 $ one obtains the family of metrics studied by Page  \cite{page1978taub}. 
For  generic values of $M$, $N$ these metrics have conical singularities while special values give the product metric on $S ^2 \times S ^2 $, a $\mathbb{Z}  _2 $ quotient of the round metric on $S ^4 $, the Fubini-Study metric on $\mathbb{C}  P ^2 $ and Page's metric on $\mathbb{C}  P ^2 \#\overline{\mathbb{C}  P }^2  $.

In this paper we are going to consider the case $\Lambda <0 $. Only two of the three parameters $\{\Lambda , M , N \} $  are effective as, up to a global rescaling of the metric,  one of them can always be absorbed by a redefinition of the coordinates, so  from now on we set 
\begin{equation}
 \Lambda =-3 .
 \end{equation} 
Since (\ref{bb9}), (\ref{deltapol}) only depend on $N ^2 $, we take $N > 0 $, which by (\ref{dvol_form}) is equivalent to fixing the orientation of the underlying manifold. Suitably taken, the limits $N \rightarrow 0 $ and $N \rightarrow \infty $ also yield interesting metrics, the Schwarzschild-anti de Sitter one in the former case, and the Eguchi-Hanson-anti de Sitter one in the latter, but we will not considered them here. The Komar mass $M$ is for now allowed to take any real value.

Asymptotically  (\ref{bb9}) approaches the metric 
\begin{equation}\label{Asympt_biaxial}
g_{r \gg 1} \simeq  \frac{\mathrm{d} r^2}{r^2}  + r^2\left(  4 N ^2  \eta _3 ^2 + d\Omega ^2  \right).
\end{equation}
The Riemann tensor of (\ref{Asympt_biaxial}) is that of a space of constant sectional curvature $-1$ up to corrections of order $O(r^{-3})$. For  $M =0 $, $N =  1/2$,  (\ref{bb9}) is exactly the metric of  $H^4$.

It is interesting to compare the asymptotic behaviour (\ref{Asympt_biaxial})  with that of the Ricci-flat case $\Lambda =0 $ \cite{franchetti2019harmonic}. If $\Lambda =0 $ the coefficients of the $ \eta _1 $ and $\eta _2 $ terms scale with $r ^2 $ while that of  $\eta_3 $  is asymptotically constant. Therefore the volume of the base of the asymptotic circle fibration grows unboundedly with $r$ while the fibres approach a fixed length. More formally, one could say that asymptotically the volume of a geodesic ball grows like the cube of its radius --- to be contrasted with the fourth power behaviour of Euclidean 4-space. For historical reasons, spaces exhibiting this behaviour are called ALF. In contrast, if $\Lambda <0 $ all the $\eta _i $ terms scale exponentially with respect to geodesic distance, and we are dealing with a family of conformally compact metrics \cite{anderson2005topics}.

\subsection{Particular cases of bi-axial Bianchi IX metrics}
The metric (\ref{bb9}), (\ref{deltapol})    is smooth only for some particular values of the parameters $M , N $ which we will now review. 
It can be checked (e.g.~by calculating the $L ^2 $-norm of the Riemann tensor) that (\ref{bb9}) has a curvature singularity at $r =  N  $ unless  (\ref{deltapol}) has a double zero there.   Hence if $\rb $ is the largest real root of $\Delta(r) $, in order to get a complete non-singular metric we have two possibilities:
\begin{itemize}
\item [ (i) ] $\rb =N $ is a double root of $\Delta(r) $ and  $r \in [ N , \infty )$.
\item [ (ii) ] $\rb > N  $ is a single root  and   $ r \in [ \rb , \infty )$.
\end{itemize} 
We do not need to consider the case of $r$ ranging from  $- \infty $ to the smallest root of $\Delta(r) $  since (\ref{bb9}) is invariant under the simultaneous change $r \rightarrow -r $, $M \rightarrow - M $.

\subsubsection{Solutions of NUT type}

Consider first the case where $\Delta (r) $ has a double root at $r = N $.
\label{tnadscase} 
Imposing 
\begin{align}
\label{cq} 
\Delta ( N)  &=0=\Delta ^\prime (N)
\end{align} 
gives $M =\Msd $ where
\begin{equation}
\label{masstn} 
 \Msd=N   \left( 1- 4 N ^2  \right).
\end{equation}
The polynomial (\ref{deltapol})  factorises as
$ (r -  N )^2   (r- \ran)(r - \rbn  ) $,
where
\begin{equation}
\label{radtn} 
\ran  =  -   N +    \sqrt{  4 N ^2 - 1}, \quad \rbn  =  -   N -    \sqrt{  4 N ^2 - 1} .
\end{equation} 
The largest root is $r =N>0 $, so we take $r \in [ N , \infty )$. The resulting metric is the Taub-NUT-AdS (TN-AdS) metric \cite{pedersen:1986},
\begin{equation}
\begin{split} 
\label{tnmetric} 
\gtn &=
\left( \frac{r ^2 - N ^2 }{\Deltatn (r)} \right) \mathrm{d}  r ^2 +  (r ^2 - N ^2 ) \mathrm{d}  \Omega ^2 +  4 N ^2 \left( \frac{\Deltatn (r)  }{r ^2 -  N ^2   }\right)  \eta _3  ^2 ,\\
\Deltatn (r) &=  (r -  N )^2   (r- \ran )(r - \rbn).
\end{split} 
\end{equation}
Note that $N$ is allowed to take any value in $(0, \infty )$.
The $+ $ subscript in $\gtn $ and other symbols  is due to the fact that the associated Weyl tensor is self-dual, a fact which will play a role in the analysis of Killing spinors of Section \ref{kgspsss}.

The coefficients of  $ \mathrm{d}  \Omega ^2 $ and $\eta _3  $  in (\ref{tnmetric}) both vanish at $r =N $ which is thus a NUT. 
Regularity at the NUT can be checked by substituting $r -N=  \rho ^2 /(8 N )$ and Taylor expanding (\ref{tnmetric})  near  $\rho =0 $, getting
\begin{equation}
\label{nearnut} 
g = \mathrm{d}  \rho ^2 +  \frac{\rho ^2}{4} \left(  \eta _1 ^2 +\eta _2 ^2 +  \eta _3 ^2\right), 
\end{equation}
which is the Euclidean metric on $\mathbb{R}  ^4 $. Hence the geometry near the NUT is that of flat space, just as in the case of ordinary Taub-NUT, which is obtained as $\Lambda \rightarrow 0 $.
TN-AdS is thus a smooth complete Einstein metric on a manifold with the topology of $\mathbb{R}  ^4 $. 
For the special value $N =  1/2 $ (\ref{masstn}) gives $M =0 $, $\Deltatn (r) $ simplifies to
$(r ^2 - \tfrac{1}{4} ) ^2$ 
and (\ref{tnmetric})  becomes the metric of hyperbolic space  $H ^4 $.

\subsubsection{Solutions of bolt type}
Suppose now the largest real root $\rb $ of (\ref{deltapol}) satisfies $\rb > N $. The coefficient of $ \mathrm{d}  \Omega ^2 $ in (\ref{bb9}) is  non-vanishing at $r _1 $, so  $\{r = \rb \}$ is a bolt with the topology of a 2-sphere. We need to ensure that the metric has no conical singularities around the bolt.

For  fixed $( \theta , \phi )$, the geometry near the bolt is that of a  2-dimensional disk parametrised by $(r, \psi )$. For a smooth metric, as $r $ approaches $r _1 $ the geometry becomes increasingly Euclidean. Therefore  the ratio $ \mathrm{d}  L / \mathrm{d}  \epsilon $, where $L$ is the length of the small curve consisting of points at geodesic distance $\epsilon$ from the 2-sphere $r = \rb $,  converges as $\epsilon \rightarrow 0 $ to its Euclidean value of $2 \pi $. In general, consider a 2-dimensional metric of the form
\begin{equation}
ds^2 = F ^2 (r) \mathrm{d}  r ^2 + C  ^2 (r)  \mathrm{d} \psi ^2, 
\end{equation} 
with a radial coordinate $r \geq \rb >0$ and an angular coordinate $\psi $ with period $T$. Then
\begin{equation*}
\mathrm{d}  \epsilon = F \, \mathrm{d}  r, \quad 
  L = \int _0 ^T C (r(\epsilon)) \mathrm{d}  \psi =T\,  C(r (\epsilon) ) ,
 \end{equation*}
hence the smoothness condition at $r =\rb $ is 
\begin{equation}
\label{boltcondition} 
2 \pi 
= T \lim _{ r \rightarrow \rb }\frac{ \mathrm{d}  C }{ \mathrm{d}  r} \frac{ \mathrm{d}  r}{ \mathrm{d} \epsilon } 
= T \lim _{ r \rightarrow \rb } \frac{C ^\prime }{F }.
\end{equation} 

Substituting in (\ref{boltcondition}) the expressions for $F$ and $C$  from (\ref{bb9}), $F = -f = \sqrt{\frac{ r ^2 - N ^2 }{\Delta (r) } }$, $C =2 N /F $, and the period $ T =4 \pi /p $ of $\psi$, we get
\begin{equation}
\label{dprcond} 
\frac{\Delta ^\prime (\rb )}{\rb ^2 - N ^2 } = \frac{p }{ 2 N  }.
\end{equation} 
Expressing $M $ in terms of $\rb $ from the condition $\Delta (\rb )=0 $, substituting in (\ref{dprcond}) and solving for $\rb $ gives two solutions,
\begin{equation}
\rho _1  = \frac{p -  \sqrt{  p ^2 -48 N ^2 + 144 N ^4  }}{12 N }, \quad \rho _2  = \frac{p +   \sqrt{  p ^2 -48 N ^2 + 144 N ^4  }}{12 N }.
\end{equation} 
The solution $\rho _1 $  satisfies $\rho _1  > N $ only for
\begin{equation} 
\label{bcm} 
p =1 \quad \text{and} \quad 0<N \leq \frac{1}{2} \sqrt{\frac{ 2- \sqrt{ 3}}{3}}.
\end{equation} 
Solving $\Delta (\rho _1 ) = 0 $ for $M$ then gives
\begin{equation}
\label{m-} 
M _1  = \frac{1 }{864 N ^3 } \left( 1 -  (1 + 24 N ^2  -288 N  ^4 )\sqrt{ 1-48 N ^2  + 144 N ^4  } \right).
\end{equation} 

The condition $ \rho _2  > N $  instead admits solutions for any positive integer $p$, 
\begin{equation}
\label{bcp} 
\begin{cases}
p =1 \quad  \text{and} \quad  & 0< N \leq \frac{1}{2} \sqrt{ \frac{2-\sqrt{ 3 }}{3}} ,\\
p =2 \quad  \text{and} \quad  & 0< N  <\frac{1}{\sqrt{ 6 }} ,\\
p \geq 3 \quad \text{and}\quad  &0< N < \infty.
\end{cases} 
\end{equation} 
The  value of $M$ corresponding to $\rho _2  $ is
\begin{equation}
\label{mpplus} 
\begin{split} 
M _2 ^p  &=
\frac{1}{864 N ^3 } \left(p ^3 +  \left(p ^2 +  24 N ^2 -288 N ^4   \right) \sqrt{ p ^2 -48 N ^2 + 144 N ^4 }    \right).
\end{split} 
\end{equation} 

These bolt-type solutions are smooth complete Einstein metric with the topology of the total space of the complex line bundle $ \mathcal{O} (-p ) \rightarrow \mathbb{C}  P ^1  $. The  case $p =1 $, with $N$ constrained by either (\ref{bcm}) or (\ref{bcp})  and with the corresponding value of $M $, is known as the Taub-bolt-AdS (TB-AdS) metric \cite{chamblin1999large}.
Reinstating the cosmological constant $\Lambda$ one notes  that $ \lim _{ \Lambda \rightarrow 0 } \rho _1  =2 N $, twice the radius of the bolt in the Ricci-flat Taub-bolt space, while $\rho _2  $ diverges in this limit. Thus, as already noted in \cite{chamblin1999large}, only the solution with $r _1 =\rho _1  $ converges  to the Ricci-flat Taub-bolt space as $\Lambda \rightarrow 0 $. 

\subsubsection{Half-conformally-flat solutions}
\label{hcfsolutions} 
In Section \ref{kgspsss} we will need to know for which values of the parameters $M $, $N $ the metric (\ref{bb9}), (\ref{deltapol})  is half-conformally-flat, that is has  self-dual or anti self-dual Weyl tensor.
It can be checked that, for the choice of orientation  (\ref{dvol_form}),  the Weyl tensor of (\ref{bb9}) is self-dual if and only if 
\begin{equation}
\label{msd} 
\Msd = N (1- 4 N ^2 ) ,
\end{equation} which gives the TN-AdS solution (\ref{tnmetric}), and  is anti self-dual if and only if  $M$ takes the opposite value
\begin{equation}
\label{masd} 
\Masd = -N (1- 4 N ^2 ) .
\end{equation} 

We write $\Deltab (r) $  for  (\ref{deltapol}) with  $M=\Masd $, 
\begin{equation}
\label{drmn} 
\Deltab (r) =  (r +  N )^2   (r- \rab )(r - \rbb),
\end{equation} 
where
\begin{equation}
\label{rpmddd} 
\rab  =  N +    \sqrt{  4 N ^2 - 1}, \quad \rbb  =  N -    \sqrt{  4 N ^2 - 1} .
\end{equation} 
and $\rab$ is the largest root.

There are no simultaneous solutions of  (\ref{msd}) or (\ref{masd}) and  (\ref{m-}) or (\ref{mpplus}) with $p =1  $, thus TB-AdS is not half-conformally-flat for any value of $N$. However for $p > 1 $ both (\ref{masd}) and (\ref{mpplus}) are satisfied if  $N =N _p $, having defined
\begin{equation}
\label{nspehcf} 
N_p  = \frac{p}{4 \sqrt{ p-1 }}, \quad  p \geq 3.
\end{equation} 
We need to take $p \geq 3 $ in (\ref{nspehcf}) in order to satisfy (\ref{bcp}).
The values of  $M$ and $r _1 $ corresponding to (\ref{nspehcf}) are
\begin{equation} 
 \label{mrspec} 
\begin{split}
 M_p  &=-\frac{p (p-2 )^2 }{16 (p-1 )^{ 3/2 }}, \quad 
(r _1)_p  =\frac{3p-4}{4 \sqrt{ p-1 }}.
\end{split} 
\end{equation} 
The  metric (\ref{bb9}), (\ref{deltapol}) with $N =N _p $, $M=M _p $ and $p \geq 3 $  is  the quaternionic Eguchi-Hanson solution \cite{lebrun1988,hitchin:1995a},
\begin{equation}
\begin{split} 
\label{qeh} 
\gtb  &=
\left( \frac{r ^2 - N_p  ^2 }{\Deltabp (r)} \right) \mathrm{d}  r ^2 +  (r ^2 - N _p ^2 ) \, \mathrm{d}  \Omega ^2 +  4 N_p  ^2 \left( \frac{\Deltabp  (r)  }{r ^2 -  N_p  ^2   }\right)  \eta _3  ^2 ,\\
\Deltabp (r) &=  \left( r +   \frac{p}{4 \sqrt{ p-1 }} \right) ^2   \left( r+  \frac{4-3p}{4 \sqrt{ p-1 }} \right)  \left( r +\frac{4-p}{4 \sqrt{ p-1 }}  \right), \quad p \geq 3 .
\end{split} 
\end{equation}
Note that the parameters in (\ref{mrspec}) are well-defined and real for $p =2 $. The corresponding metric is that of $H ^4/ \mathbb{Z}  _2  $ which is conformally flat, but  has a conical singularity at $r =0 $ because of the $\mathbb{Z}  _2 $ quotient. As we have already seen, the smooth $H ^4 $ metric can be obtained for $p =1 $ as a special case of (\ref{tnmetric}). 

To summarise, for  $M$ given by (\ref{msd}) we have the 1-parameter family (\ref{tnmetric}) of self-dual solutions of NUT type. For $M $ given by  (\ref{masd})  we have the infinite discrete family (\ref{qeh}) of anti self-dual solutions of bolt type. The self-dual and anti self-dual branches join at $M =0 $, which gives the conformally flat metric of $H ^4 $.

\section{Spherically symmetric harmonic $2$-forms}
\label{harmonicforms} 
We are going to study the harmonic cohomology, that is the space of harmonic square-integrable differential forms, of the Riemannian manifold $\mathcal{M}$ equipped with the metric (\ref{bb9}), (\ref{deltapol}). We recall that depending on the values of the parameters $M$, $N$ and $p$, $\mathcal{M}$ is either diffeomorphic to $\mathbb{R}  ^4 $ or to the total space of the line bundle $ \mathcal{O} (-p )\rightarrow \mathbb{C}  P ^1  $.
We denote by $L ^2 (\mathcal{M} )$ the space of square-integrable  forms on $\mathcal{M} $. Since $\mathcal{M}$ is complete, the harmonic forms in $L^2(\mathcal{M})$ are precisely the closed and co-closed forms in $L^2(\mathcal{M})$. 

Since $\mathcal{M}$ is conformally  compact,  harmonic cohomology is trivial except in middle dimension where it is infinite dimensional \cite{mazzeo:1988}.   We will restrict our analysis to harmonic 2-forms in $L ^2 (\mathcal{M} )$  which are invariant under the $SU (2) $ subgroup of the isometry group of $\mathcal{M} $. For $\beta \in \Omega ^2 (\mathcal{M} )$ this invariance conditions reads
\begin{equation}
\label{invcond} 
L _{ Z _i } \beta =0, \quad  i =1,2,3,
\end{equation} 
where $\{ Z _i \} $ are the right-invariant vector fields on $SU (2) $,
\begin{equation}
\label{rightnvvecs} 
\begin{split}
Z _1 &= - \sin \phi\,  \partial _\theta - \frac{\cos \phi }{\sin \theta } \left( \cos \theta \, \partial _\phi - \partial _\psi \right) ,\\
Z _2 &= \cos \phi\,  \partial _\theta -  \frac{\sin  \phi }{\sin \theta } \left( \cos \theta \, \partial _\phi - \partial _\psi \right) ,\\
Z _3 &= \partial _\phi.
\end{split}
\end{equation} 

If $\beta$ is harmonic then it is closed so locally we can write
\begin{equation}
\beta =\mathrm{d}  \alpha
\end{equation}
 for some $1$-form $\alpha$.  
 Since $ L_{ Z _i} \eta _j =0$,  (\ref{invcond}) is solved by taking
\begin{equation}
\label{invariant_1form}
\alpha = \alpha _1(r) \eta_1 + \alpha _2(r) \eta_2 + \alpha _3(r)  \eta_3 ,
\end{equation} 
where the $\alpha _i $s are arbitrary functions of $r$ only. We write
\begin{equation}
\beta _i =\mathrm{d} (\alpha _i \, \eta _i ).
\end{equation} 
We could also include an exact term  $ \alpha _4(r)  \mathrm{d} r $ in (\ref{invariant_1form}), but since we are only interested in $\beta = \mathrm{d}  \alpha$ we can safely drop it. Observe that if $\beta$ is also required to be invariant   under the  $U (1) $-isometry generated by  $\partial_\psi$ then necessarily $\alpha _1 =\alpha _2 =0 $.

In terms of (\ref{invariant_1form}), the co-closure condition $\mathrm{d} ^\ast \beta =0 $ becomes
\begin{align}
\label{har1} 
&\alpha_i ^{ \prime \prime } -  \frac{2f ^\prime }{f}  \alpha_i ^\prime - \frac{  f ^4}{4 N ^2 } \alpha_i =0,\quad  i  =1 ,2, \\
\label{har3} 
&\alpha_3 ^{ \prime \prime } +   \frac{2a ^\prime }{a} \alpha_3 ^\prime  - \frac{4 N ^2  }{a ^4 } \alpha_3 =0.
\end{align}

Equation (\ref{har3}) can be integrated explicitly, obtaining
\begin{equation}
\label{alpha3} 
\alpha _3 = C _1  \left( \frac{r ^2 + N ^2 }{r ^2 - N ^2 }\right) - C _2  \left( \frac{2 N r}{r ^2 - N ^2 } \right) ,
\end{equation} 
with $C _1 $, $C _2 $ arbitrary constants.
The  form  $\beta _3 = \mathrm{d}  (\alpha _3 \eta _3 )$ is given by
\begin{equation}
\label{frcurv} 
\begin{split} 
\beta _3 = - \frac{1}{(r ^2 - N ^2 )^2 } \left[  \Big(C _1 (r ^2 + N ^2 ) - 2 C _2 r N \Big) e ^1 \wedge e ^2 +  \Big(C _2 (r ^2 + N ^2 )- 2 C _1 r N  \Big)e ^3 \wedge e ^4  \right].
\end{split} 
\end{equation} 
Clearly (\ref{frcurv}) is self-dual (respectively anti self-dual) if and only if $C _1 =C _2   $ ($C _1 =- C _2 $).  We denote by 
$\beta _3 ^+ $,  the self-dual form obtained by setting $C _1 =C _2 =-1 $,
\begin{equation}
\label{beta3plus} 
\beta _3 ^+ =  \frac{ 1}{(r +  N  )^2 } ( e ^1 \wedge e ^2 + e ^3 \wedge e ^4 ).
\end{equation} 
and by $\beta _3 ^- $  the anti-self-dual form obtained by setting $C _1 =-C _2 =-1 $,
\begin{equation}
\label{beta3minus} 
\beta _3 ^- =  \frac{ 1}{(r -  N  )^2 } ( e ^1 \wedge e ^2 - e ^3 \wedge e ^4 ).
\end{equation} 
The form $\beta _3 $ can then be written as
\begin{equation}
\label{b3deco} 
\beta _3 = \left( \frac{C _1 + C _2 }{2} \right) \beta _3 ^+ + \left( \frac{C _1 - C _2 }{2} \right) \beta _3 ^- .
\end{equation} 
We also define
\begin{equation}
\alpha _3 ^\pm = \left( \frac{r- N }{r + N } \right) ^{ \pm 1 }
\end{equation} 
so that $\beta _3 ^\pm = \mathrm{d} (\alpha _3 ^\pm \eta _3 )$.

Consider now the ODE (\ref{har1}). Taking the ansatz
\begin{equation}
\alpha _i (r)  
= \exp ( l  (r) ), 
\end{equation} 
 and substituting in  (\ref{har1}) one finds the condition
\begin{equation}
\label{fhicond} 
l ^\prime = \pm  \frac{1 }{2 N } f ^2 = \pm  \frac{1 }{2 N }\left(  \frac{r ^2 - N ^2 }{\Delta(r)  }\right) .
\end{equation} 
Defining
\begin{equation}
\label{phiexpr} 
\varphi (r) =-\frac{1 }{2 N } \int  \left(  \frac{r ^2 - N ^2 }{\Delta(r)  }\right)\mathrm{d} r,
\end{equation} 
the general solution of (\ref{har1}) is 
\begin{equation}
\label{kjjsol} 
\alpha _i (r)  
= \left( \frac{K _i - L _i }{2} \right) \exp (\varphi (r) )+  \left( \frac{K _i +  L _i  }{2} \right) \exp (-\varphi (r) ), 
\end{equation} 
with $K _1 , K _2, L _1 , L _2  $ arbitrary constants. 
For $\alpha_i $ given by (\ref{kjjsol}),  $\beta _i $  is self-dual (respectively anti self-dual) if and only if $K _i =L _i    $ ($K _i =- L _i  $). Thus  defining
\begin{equation}
\label{alphaplusdef} 
\alpha _i ^\pm = \exp(\mp  \varphi ), \quad 
\beta  _i ^\pm = \mathrm{d} \left(\alpha _i ^\pm  \eta _i \right) , \quad i =1,2,
\end{equation} 
we have
\begin{equation}
\label{alphaiexprrr} 
\alpha _i = \left( \frac{K _i - L _i }{2} \right)  \alpha _i ^- +  \left( \frac{K _i +  L _i  }{2} \right) \alpha _i ^+ 
\end{equation} 
and similarly for $\beta _i $.

The precise form of $\varphi$  depends on whether any of the roots $\{ r _1, r _2 , r _3 , r _4  \} $ of $\Delta (r) $ are repeated. We  will write down an explicit expression  in the case of TN-AdS, where  $r _1 =r _2 =N $ and $r _3 $, $r _4 $ are distinct, see Section \ref{hartn},  and in  the generic case for bolt-type solutions, where $\{ r _i \} $ are all distinct, see Section \ref{h2ftbads}. The other cases can be handled similarly.

In order to check the $L ^2 $ condition we will make use of the following expression, valid  for any exact 2-form $\beta  = \mathrm{d} \alpha $ on a dense subset  of a  space with metric (\ref{bb9}), 
\begin{equation}
\label{l2form} 
\mathcal{\beta } \wedge * \mathcal{\beta } 
= \left[ \left( \frac{ \alpha   ^\prime  } {f  }\right) ^2  + \left( \frac{\alpha   f}{2 N } \right) ^2  \right]  \frac{ \operatorname{vol }}{ a ^2 }.
\end{equation} 
In particular, the form (\ref{frcurv}) has norm
\begin{equation}
\label{f3case} 
\| \beta_3\|^2 =   \int _{\mathcal{M}} \beta _3 \wedge * \beta _3 
= 8 \pi \, T  N \lim _{ r \rightarrow \rb}   \frac{r}{(r ^2 - N ^2 ) ^2 }  \Big( (C _1 ^2 + C _2 ^2 ) (r ^2 + N ^2 ) -4 C _1 C _2 N r \Big),
\end{equation} 
for $r _1 $ the largest root of $\Delta(r) $ and $T$ the period of $\psi$.

\subsection{Spherically symmetric harmonic 2-forms on solutions of NUT type}
\label{hartn} 
Let $\mathcal{M}$  be TN-AdS. Consider $\beta _3 $ first.  Since $r _1 =N $, by (\ref{frcurv})    $\beta _3  $ is well defined at $r =N $ if and only if it is self-dual, so we need to take
\begin{equation}
\alpha _3 = C \alpha _3 ^+ . 
 \end{equation}
The form $\beta _3 ^+ $  is in $L ^2 (\mathcal{M} )$ as, by (\ref{f3case}) with $T =4 \pi $, $r _1 =N $,
\begin{equation}
\| \beta _3 ^+   \| ^2   =  16 \pi ^2  .
\end{equation} 
Since $\alpha _3 ^+ (N ) =0 $, $\alpha ^+ _3 \eta _3 $ is a globally defined 1-form. Thus $\beta _3 ^+   $ is exact, which had to be the case since TN-AdS is topologically trivial.

Consider now the forms $\beta _i  $ for $i =1,2 $.
 $\Deltatn (r)   $ has a double zero at $r = N $ and  factorises as
\begin{equation}
\label{rpluminus} 
\Deltatn  (r) = (r- N )^2 (r- \ran )(r - \rbn ), \quad 
\ran  = - N +   \sqrt{4 N ^2 - 1  }, \quad \rbn = - N -  \sqrt{4 N ^2 - 1  }.
\end{equation} 
The integral in (\ref{phiexpr}) thus gives
\begin{equation} 
\label{varphi_TN}
 \varphitn (r) =\frac{N ^2 - \ran ^2 }{2 N\Deltatn^\prime  (\ran )} \log( r - \ran )
+ \frac{N ^2 - \rbn ^2 }{2 N\Deltatn^\prime  (\rbn )} \log (r - \rbn) -   \log( r-N).
\end{equation} 

Suppose $\ran =\rbn $ first, so that we are dealing with hyperbolic 4-space. Then $ \varphitn=- \log (r- N )$ and
\begin{equation}
\alpha_i  = \left( \frac{K _i -L _i  }{2}  \right) \frac{1}{r-N} + \left( \frac{K _i + L _i  }{2} \right) (r- N).
\end{equation} 
It is easily checked that  $\beta _i $ does not belong to $L ^2(\mathcal{M}) $.

Assume now $\ran \neq \rbn$.
From (\ref{alphaplusdef}), (\ref{varphi_TN})  we see that  $\lim _{ r \rightarrow N  ^+ } \alpha _i ^+ =0$,  $\lim _{ r \rightarrow N  ^+ } \alpha _i ^- =\infty $. Thus to have an $L ^2 $ form we need to take
\begin{equation} 
\alpha _i = K _i \alpha _i ^+ .
\end{equation} Asymptotically, to leading order,
\begin{equation}
\alpha ^+ _i  (r) =   \mathrm{e} ^{ - \varphitn(r)  }
\simeq  r ^{ 
-\frac{1}{2 N } \left(\frac{N ^2 - \ran^2 }{\Deltatn ^\prime (\ran)} +  \frac{N ^2 - \rbn ^2 }{\Deltatn ^\prime (\rbn) }-2N\right)  }
 =1 ,
\end{equation} 
having used
\begin{equation}
\frac{ N ^2 - \ran^2 }{\Deltatn ^\prime (\ran )} +  \frac{N ^2 - \rbn^2 }{\Deltatn ^\prime (\rbn) } -2N=0.
\end{equation} 
We can calculate the $L ^2 $-norm using Stokes' theorem,
\begin{equation}
\label{sol3} 
\| \beta ^+   _i \| ^2 
= - \int _{ \partial M }(\alpha ^+  _i)  ^2\,   \eta _1 \wedge  \eta _2 \wedge \eta _3  
= 16 \pi ^2  \lim _{r \rightarrow \infty} ( \alpha _i ^+ )  ^2 =16 \pi ^2 .
\end{equation}  
Both $\beta _1 ^+ $ and $\beta _2 ^+ $ are globally defined exact 2-forms. It can be checked that none of  $\{ \beta _1^+  , \beta _2 ^+  , \beta _3 ^+  \} $ is $L ^2 $-exact.

Summarising,  the space of  square-integrable $SU (2) $-invariant  harmonic 2-forms on TN-AdS consists entirely of self-dual exact but not $L ^2 $-exact  forms. For  $ N\in (0, 1/2 )\cup (1/2 , \infty )  $ it is 3-dimensional and spanned by  $(\beta _1 ^+ , \beta _2 ^+ , \beta _3 ^+ )$, which for TN-AdS take the form
\begin{equation} 
\label{tnhforms} 
\begin{split} 
\btnone&= \mathrm{d} \left[ \mathrm{e} ^{ - \varphitn }\eta _1  \right] , \quad 
\btntwo= \mathrm{d} \left[ \mathrm{e} ^{ - \varphitn }\eta _2  \right] , \quad 
\btnthree= \mathrm{d} \left[    \left( \frac{r- N  }{r + N }\right)  \eta _3  \right],
\end{split}
\end{equation} 
with $\varphitn $ given by (\ref{varphi_TN}).
For $ N  = 1/2 $ we obtain $H ^4 $ and the space  of  square-integrable $SU (2) $-invariant  harmonic 2-forms is 1-dimensional and spanned by $\btnthree $ only.

\subsection{Spherically symmetric harmonic 2-forms on solutions of bolt type}
\label{h2ftbads} 

Let $\mathcal{M}$  be a solution of bolt type. The form $\beta _3 $ is again given by (\ref{frcurv}).
This time since $r _1  > N $,  (\ref{f3case}) 
is finite for any value of $C _1 $, $C _2 $ so $\beta _3 \in L^2(\mathcal{M})$.  The 2-form $ \beta _3 $ is globally defined, however since $\psi $ is not defined at the bolt $\{r =r _1 \}$, the 1-form $\alpha _3 \eta _3 $ is not globally defined unless  $\alpha _3  (r _1 ) =0 $.
We can write
\begin{equation}
 \alpha _3 = \tilde C _1  \left( \frac{2 N r _1 }{r _1 ^2 + N ^2 }\,   \frac{r ^2 + N ^2 }{r ^2 - N ^2 } 
 - \frac{2 N r}{r ^2 - N ^2 } \right) 
 + \tilde C _2  \left( \frac{2 N r _1 }{r _1 ^2 + N ^2 }\,   \frac{r ^2 + N ^2 }{r ^2 - N ^2 } 
+  \frac{2 N r}{r ^2 - N ^2 } \right) ,
\end{equation} 
with 
\begin{equation}
\tilde C _1 =  \frac{C _2 }{2} + \frac{C _1 }{2}  \left( \frac{r _1 ^2 + N ^2 }{2 N r _1  }\right) , \quad 
\tilde C _2 = - \frac{C _2 }{2} + \frac{C _1 }{2}  \left( \frac{r _1 ^2 + N ^2 }{2 N r _1  }\right) .
\end{equation} 
We thus have $\alpha _3 (r _1 )=0 $ if and only if $\tilde  C _2 =0$, or equivalently 
\begin{equation}
C _2 =\left( \frac{r _1 ^2 + N ^2 }{2 N r _1 } \right)  C _1 .
\end{equation} 
The 1-dimensional subspace of $\{ \beta _3: (\tilde C _1, \tilde C _2)  \in \mathbb{R}  ^2 \} $ obtained by setting $\tilde C _2 = 0 $ is exact.  The complementary subspace obtained by setting $\tilde C _1 =0 $ generates $ H ^2 _{ \text{dR} }( \mathcal{M}  )\simeq \mathbb{R}  $. Since $r _1 > N $, the conditions $ C _2 = \pm  C _1 $, $C _2 =\pm  \frac{r _1 ^2 + N ^2 }{2 N r _1  } C _1 $ are incompatible for any choice of signs; in particular  if $ \beta _3 \neq 0 $ then it cannot be (anti) self-dual and exact. In fact writing $\beta _3 = \tilde C _1 \beta _3 ^{ \mathrm{exact}} +\tilde C _2   \beta _3 ^{ H ^2} $ we find
\begin{equation}
\begin{split} 
\be & =\frac{1 }{4N r _1} \Big( (r _1 + N )^2  \beta _3 ^+ - (r _1 - N )^2 \beta _3 ^- \Big),\\
\beta _3 ^{ H ^2 }  &=\frac{1}{4N r _1} \Big( (r _1 + N )^2  \beta _3 ^- - (r _1 - N )^2 \beta _3 ^+  \Big).
\end{split} 
\end{equation}

The form $\be $   is exact but not $ L ^2 $-exact. In fact, $ \be = \mathrm{d} (\ae\eta _3  + \gamma )$ where $\ae $ is (\ref{alpha3}) with $\tilde C _2 =0 $ and $\gamma$ is any closed 1-form. Since $H ^1 _{ \mathrm{dR} } (\mathcal{M} ) = 0 $ we can write $ \gamma = \mathrm{d} f _3 $ for some smooth function $f _3 $. Then
\begin{equation}
\label{3l2e} 
\| \ae \eta _3 + \mathrm{d} f _3  \| ^2 = \| ( \ae ) ^2 \eta _3 \| ^2 + \| \mathrm{d} f _3 \| ^2 + 2  \langle \ae \eta  _3 , \mathrm{d} f _3 \rangle .
\end{equation} 
We have
\begin{equation} 
 \| (\ae ) ^2 \eta  _3 \| ^2 = 16 \pi^2  N  \int _{ r _1 }^\infty \frac{ (r ^2 - N ^2 ) ^2 }{\Delta(r)  }\, (\ae) ^2 \,  \mathrm{d} r,
\end{equation} 
which diverges since the integrand is asymptotically a non-zero constant.  Hence in order for (\ref{3l2e}) to be finite we would need $\mathrm{d} f _3 $ to be asymptotically equal to $- \ae \eta _3 $, which is not possible as the latter is not asymptotically exact.

Let 
$\{r _1 , r _2 , r _3, r_4 \} $ be the roots of the polynomial $\Delta (r) $, labelled so that $\rb>N $ is the largest positive root, and assume they are all distinct.
Integrating (\ref{fhicond}) then gives
\begin{equation}
\label{varfi} 
\varphitb (r) = \frac{1}{2 N } \sum _{i  =1  }^4 \frac{ N ^2 - r_i ^2 }{\Delta ^\prime (r_i)} \log (r- r_i ) .
\end{equation} 
It is convenient at this point to  recall that if 
\begin{equation}
p (r) =\prod _{ i =1 }^n (r - r _i )
\end{equation}
is a monic polynomial of degree $n \geq 4$ with distinct roots $r _i $ then
\begin{align}
\label{idp1} 
\sum _{ i =1 }^n &\frac{1}{p ^\prime (r _i )} =0,\\
\label{idp2} 
\sum _{ i =1 }^n &\frac{r _i }{p ^\prime (r _i )} =0,\\
\label{idp3} 
\sum _{ i =1 }^n &\frac{r _i ^2 }{p ^\prime (r _i )} =0.
\end{align}
These identities  follow from the more general one, valid for any function $h$,
\begin{equation}
\sum _{ i = 1 }^k   \frac{h (r _i )}{p ^\prime (r _i )} =h [r _1 , \ldots , r _k ],
\end{equation} 
with $h [ r _1 , \ldots , r _k ] $ the $k$-th divided difference of $h$. Taking $h (x) =1 $, $k \geq 2 $ gives (\ref{idp1});  taking $h (x) =x $, $k \geq 3 $ gives (\ref{idp2}) and taking $h (x) =x ^2 $, $k \geq 4 $ gives (\ref{idp3}).

We now check square-integrability. Since $r _1 >N >0 $ and $\Delta ^\prime (r _1 )> 0 $,
\begin{equation}
\lim _{ r \rightarrow r _1 ^+ }\varphitb(r)  = +  \infty. 
\end{equation}
Therefore by (\ref{kjjsol}), $\alpha_i ^2$ is unbounded in a neighbourhood of the bolt $\{r =r _1 \} $ unless $K _i = L _i $ so that  $ \alpha _i =K _i \alpha _i ^+ $. 
Since $\lim _{ r \rightarrow r _1 } (\alpha _i ^+ )^2   =0 $, we obtain
\begin{equation}
\label{sol3} 
\| \beta  _1 ^+  \| ^2 
= - \int _{ \partial \mathcal{M} } (\alpha _i ^+ ) ^2  \eta _1 \wedge  \eta _2 \wedge \eta _3 
 = 4 \pi T   \lim _{r \rightarrow \infty}  (\alpha _i ^+ )  ^2 .
\end{equation}  
 To evaluate the limit at infinity note that by (\ref{idp1}), (\ref{idp3}), to leading order,
\begin{equation}
\alpha_i ^+   \simeq   r ^{  \frac{1}{2 N }\sum _{ i =1 }^4 \frac{N ^2 - r _i ^2 }{ \Delta ^\prime (r _i )}} =1 .
\end{equation} 
Therefore,
\begin{equation}
\|\beta  _1 ^+  \| ^2 =\|\beta  _2 ^+  \| ^2 = \frac{ 16 \pi ^2 }{p} 
\end{equation} 
having substituted $T =4 \pi /p $ for the period of $\psi$.
Since $\alpha _i ^+ (\rb )=0 $,  the 1-form $\alpha_i ^+  \eta _i $, $i =1,2 $, is globally defined and $\beta _i ^+  $ is exact, but, essentially by the same argument used for $\beta _3  ^+ $, not $L ^2 $-exact. 

To summarise,  the space of square-integrable $SU (2) $-invariant harmonic 2-forms on bolt-type solutions is 4-dimensional  and spanned by  $(\beta _1 ^+, \beta _2 ^+ , \beta _3 ^+ , \beta _3 ^- )$, which for bolt-type solutions take the form
\begin{equation} 
\label{bforms} 
\begin{split} 
\bbone&= \mathrm{d} \left[ \mathrm{e} ^{ - \varphitb }\eta _1  \right] , \quad 
\bbtwo= \mathrm{d} \left[ \mathrm{e} ^{ - \varphitb }\eta _2  \right] , \\
\bbthreep &= \mathrm{d} \left[    \left( \frac{r- N  }{r + N }\right)  \eta _3  \right], \quad 
\bbthreem  = \mathrm{d} \left[    \left( \frac{r + N  }{r - N }\right)  \eta _3  \right],
\end{split}
\end{equation} 
with $\varphitb $ given by (\ref{varfi}).
The 3-dimensional subspace spanned by $(\bbone,\bbtwo, \bbthreep)$ comprises of self-dual forms. The 2-space spanned by $(\bbone,\bbtwo)$ is exact but not $L ^2 $-exact. The 2-space spanned by $(\bbthreep, \bbthreem ) $ can  be split into a 1-dimensional space of exact but not $L ^2 $-exact forms and a cohomologically non-trivial 1-space generating  $ H ^2 _{ \mathrm{dR}  }(\mathcal{M} ) = \mathbb{R}  $.

\section{Killing spinors}
\label{kgspsss} 
A triple $( \mathcal{M}, g, F )$, for $(\mathcal{M}, g )$ a Riemannian manifold, $F$ the curvature of a $U (1) $ connection on $\mathcal{M} $, solves the Einstein-Maxwell system with non-zero cosmological constant $\Lambda$  if
\begin{align}
\label{max} 
\mathrm{d} * F &=0,\\
\label{ein} 
R _{ \mu \nu } - \Lambda  g _{ \mu \nu } &= 2 \left( F _{ \mu \rho }F _{ \nu \sigma }g ^{ \rho \sigma } - \frac{1}{4} 	F _{ \alpha \beta }F ^{ \alpha \beta } g _{ \mu \nu } \right) .
\end{align}
The 2-form $F$ describing the electromagnetic field is thus a closed and coclosed form of the type studied in Section \ref{harmonicforms}. The condition $F \in L ^2 (\mathcal{M} )$, which physically corresponds to finite electromagnetic energy, is often also required but we will not impose it here.
A generic solution of the Maxwell equations (\ref{max}) on an Einstein manifold $ (\mathcal{M} , g )$ in general will not solve (\ref{ein}). However, if $F$ is self-dual or anti self-dual then the right hand side of (\ref{ein}) vanishes so $ (\mathcal{M} ,g, F  )$ is a solution of the coupled Einstein-Maxwell system.

Solutions of  (\ref{max})--(\ref{ein})  coincide with solutions of the bosonic sector of $4D$, $\mathcal{N} =2 $ minimal gauged  supergravity.  Supersymmetric solutions, that is admitting non-trivial Killing spinors,  are of particular interest. Killing spinors are solutions of a certain first-order equation known as the Killing spinor equation,  and the dimension of the space of solutions divided by the rank of the spinor bundle is known as the fraction of supersymmetry preserved. With respect to a local orthonormal frame $(e _\mu )$ and gauge potential $A$, the Killing spinor equation charged by the gauge potential $A$  takes the form
\begin{equation}
\label{kilspin} 
\left( \nabla _{e _\mu}  - i A (e _\mu ) I _4 + \frac{1}{2}  \Gamma _\mu  + \frac{i}{4} F _{ \nu \rho } \Gamma ^{ \nu \rho }\Gamma  _\mu  \right) \epsilon =0.
\end{equation} 
Here $\epsilon$ is a 4-component spinor field, the curvature $F$ of the $U (1) $ connection satisfies $F =\mathrm{d} A  $ and the matrices $\{ \Gamma _\mu \} $ generate a 4-dimensional representation of the Euclidean Clifford algebra $\mathrm{Cl} (0,4 ) $. We follow the convention
\begin{equation}
\Gamma _\mu \Gamma _\nu + \Gamma _\nu \Gamma _\mu = 2 \delta _{ \mu \nu },
\end{equation} 
so that
\begin{equation}
\nabla _{ e _\mu }(\epsilon) = e _\mu (\epsilon) +  \frac{1}{4} \omega _{ \rho  \sigma } (e _\mu ) \Gamma ^\rho \Gamma ^\sigma \epsilon 
\end{equation} 
for $\omega  _{ \rho  \sigma }$ the Levi-Civita connection local 1-form associated to the frame $(e _\mu )$. 

We are going to determine for which values of the parameters $M$, $N$ and choice of an (anti) self-dual curvature 2-form $F$ the metric (\ref{bb9}), (\ref{deltapol})  admits Killing spinors. This problem has been studied in \cite{martelli:2013} 
for general solutions of the Einstein-Maxwell system (\ref{max})--(\ref{ein}) with $H ^4 $ asymptotic and $SU (2) \times U (1) $ symmetry. The class of metrics studied in \cite{martelli:2013} is thus broader than (\ref{bb9}), (\ref{deltapol})  as it allows for non-Einstein manifolds. Explicitly $g$ is still given by (\ref{bb9}) but with (\ref{deltapol}) replaced by
\begin{equation}
\label{deltamar} 
\begin{split} 
 -\frac{ \Lambda }{3 } r ^4+ (1 + 2 \Lambda N ^2 )r ^2  - 2 M r + N ^2 (1 + \Lambda N ^2 ) + P ^2 - Q ^2 .
\end{split} 
\end{equation} 
The polynomial (\ref{deltamar})  differs from (\ref{deltapol})  by the addition of the constant term $P ^2 - Q ^2 $. The resulting metric is Einstein if and only if $P = \pm Q $.
The harmonic 2-form $F$ considered in \cite{martelli:2013} is
\begin{equation} 
\label{martelli2f} 
\left( \frac{ P + Q }{2} \right)  \beta _3 ^+  +  \left( \frac{P- Q  }{2} \right) \beta  _3 ^-,
\end{equation} 
where $\beta _3 ^+ $, $\beta _3 ^-  $ are given by  (\ref{beta3plus}), (\ref{beta3minus}) and  $P$,  $Q$ are the same constants as those appearing in (\ref{deltamar}).
The 2-form (\ref{martelli2f}) is the most general $SU (2) \times U (1) $-invariant harmonic 2-form on $\mathcal{M}$, but,  as shown in Section \ref{harmonicforms}, if we relax the requirement to $SU (2) $-invariance only other possibilities arise. We are thus going to study the problem for a narrower class of metrics but a broader class of 2-forms. Unfortunately, we will conclude that considering a wider class of harmonic forms leads to no solutions other than those found in  \cite{martelli:2013}.

Our search is immediately constrained by the following result \cite{dunajski:2010a}.
Let  $(\mathcal{M} , g )$ be a Riemannian 4-manifold,  $F$ a self-dual (respectively anti self-dual) harmonic 2-form. Then in order for the Killing spinor equation  (\ref{kilspin}) to admit non-trivial solutions,  $(\mathcal{M} ,g )$  must be Einstein with self-dual (respectively anti self-dual) Weyl tensor. 
The self-dual case is realised by the TN-AdS metric (\ref{tnmetric}) with  $F$ a linear combination of $\{ \beta _1 ^+, \beta _2 ^+ , \beta _3 ^+    \} $. 
The anti self-dual case by the bolt-type metric (\ref{qeh}) with $F$  a linear combination of  $\{ \beta _1 ^-, \beta _2 ^- , \beta _3 ^- \} $. Thus we will consider the form
\begin{equation}
\label{ourgenf} 
F = P \beta _3 ^{ \pm }+ K _1 \beta _1 ^\pm + K _2 \beta _2 ^\pm
\end{equation} 
with $P, K _1, K _2 $ arbitrary constants.  Locally $F =\mathrm{d} A $ with
\begin{equation} 
\label{ourgena} 
A =
  P   \left( \frac{r- N  }{r  +  N  }\right) ^{ \pm 1 }\eta _3 + \mathrm{e} ^{\mp  \varphi (r) } (K _1 \eta _1 + K _2 \eta _2 )  
\end{equation} 
and $\varphi$ given by (\ref{phiexpr}).  
The sign choice in (\ref{ourgenf}), (\ref{ourgena})   corresponds to $F =\pm * F $ with the upper sign relevant for TN-AdS and the lower sign for the anti self-dual bolt-type solutions. For $K _1 =K _2 =0 $ (\ref{ourgenf})  is equal to (\ref{martelli2f}) with $Q =\pm P $.

Before attacking the local equation (\ref{kilspin}), let us pause to consider  some global issues.
In order to admit spinors, let alone Killing ones, the manifold $\mathcal{M}$ must be spin. This is the case for TN-AdS, which is topologically trivial. Bolt-type solutions instead have the topology of the complex line bundle $\mathcal{O} (-p )\rightarrow \mathbb{C}  P ^1 $ which is spin only for $p$ even. If $\mathcal{M}$ is not spin we can equip it  with  a $\mathrm{Spin} ^{ \mathbb{C}  }$ structure by tensoring the spinor bundle on $\mathcal{M}$ with a complex line bundle $L$ having half-integer Chern number. Separately, neither the spinor bundle on $\mathcal{M}$  nor $L$  are globally well-defined, but their tensor product is \cite{lawson:1989}. Sections of the bundle so constructed give globally defined $\mathrm{Spin} ^{ \mathbb{C}  }$ spinors which are charged under the gauge potential $A$  of a connection on $L$.  If $\mathcal{M}$ is spin we obtain charged spinors by tensoring the spinor bundle with a line bundle $L$, which in this case has an integer Chern number. 

In order for  $F$ to be the curvature of a connection on the relevant line bundle $L\rightarrow \mathcal{M}$, it has to satisfy certain quantisation conditions. Since TN-AdS is topologically trivial, no conditions arise. If $\mathcal{M}$ is a bolt-type solution with even $p$, then it is spin, $L$ has integer Chern number and $F$ has to satisfy the familiar quantisation condition
\begin{equation}
\label{condtbeven} 
\frac{1}{2 \pi } \int _{ \mathrm{bolt} } F  =k\in \mathbb{Z} .
\end{equation} 
If $\mathcal{M}$ is a bolt-type solution with odd $p$ then $L$ has half-integer Chern number and $F$ has to satisfy the quantisation condition
\begin{equation}
\label{condtbodd} 
 \frac{1}{2 \pi } \int _{ \mathrm{bolt} } F  =k  + \frac{1}{2}, \quad k \in \mathbb{Z} .
\end{equation}

Let $F$ be given by (\ref{ourgenf}) and $\mathcal{M}$ be a solution of bolt type. Being exact, the forms $\beta _1 ^- $, $\beta _2 ^- $  integrate to 0 so
\begin{equation} 
 \frac{1}{2 \pi } \int _{\mathrm{bolt} } F 
= \frac{P}{2 \pi }\int _{\mathrm{bolt} } \beta _3 ^- 
= 2 P \left( \frac{r _1 + N }{r _1 - N }\right) .
\end{equation} 
Substituting (\ref{nspehcf}), (\ref{mrspec}) for $N$, $r _1 $ we thus get the condition
\begin{equation}
\label{sdqcond} 
4 P \left( \frac{p-1}{p-2} \right) =k  +\frac{1}{2} \left( p \operatorname{mod} 2 \right).
\end{equation}

We finally proceed to study (\ref{kilspin}) for $A$ given by (\ref{ourgena}), where $K _1 $, $K _2 $ are arbitrary constants and $P$ is arbitrary in the case of TN-AdS and constrained by (\ref{sdqcond}) for the anti self-dual bolt-type solutions. We mostly follow the treatment in  \cite{martelli:2013}. We already know that Killing spinors can only arise in the half-conformally-flat cases (\ref{tnmetric}), (\ref{qeh}),  with $M= \Msd$ in the former case and $M =\Masd$ in the latter,  but  since it does not require much additional effort we carry out the computation without imposing any condition on $M$, $N$.

The Killing spinor equation  can be viewed as a parallelism condition with respect to the super-covariant derivative
\begin{equation}
\mathcal{D} _{ e _\mu }  =\nabla _{e _\mu}  - i A (e _\mu ) + \frac{1}{2}  \Gamma _\mu  + \frac{i}{4} F _{ \nu \rho } \Gamma ^{ \nu \rho }\Gamma  _\mu  .
\end{equation} 
A necessary condition for the integrability of (\ref{kilspin}) is thus
\begin{equation}
\label{detcond} 
\mathcal{I} _{ \mu \nu }\epsilon =[ \mathcal{D} _\mu , \mathcal{D} _\nu ] \epsilon =0.
\end{equation} 
One calculates
\begin{equation}
\begin{split} 
\mathcal{I} _{ \mu \nu } &= \frac{1}{4} R _{ \mu \nu } ^{ \phantom{ \mu \nu }\alpha \beta } \Gamma _{ \alpha \beta } + \frac{1}{2} \Gamma _{ \mu \nu } -i F_{ \mu \nu } I _4  
+  \frac{i}{2} \nabla  _{ [ \mu } F _{ |\rho \sigma| } \Gamma  ^{ \rho \sigma } \Gamma _{\nu ]}
+  \frac{i}{4} F _{ \rho \sigma }\Gamma   _{ [ \mu }  \Gamma  ^{ \rho \sigma } \Gamma _{\nu ]}\\ &
- \frac{1}{16} F _{ \alpha \beta }F _{ \rho \sigma }[\Gamma ^{ \alpha \beta } \Gamma _\mu ,\Gamma ^{ \rho\sigma  } \Gamma _\nu ] 
+ \frac{i}{4} F _{ \rho \sigma }\Gamma ^{ \rho \sigma } \Gamma _{ \mu \nu }.
\end{split} 
\end{equation}

In order for (\ref{detcond}) to admit non-trivial solutions 
\begin{equation}
\label{deti0} 
\det (\mathcal{I} _{\mu \nu } )=0 
\end{equation} 
needs to hold for all values of $\mu$, $\nu$. Taking the gamma matrices in the chiral representation
\begin{equation}
\label{gammarep} 
\Gamma _i = \begin{pmatrix}
0 &\sigma _i  \\
\sigma _i  &0
\end{pmatrix} , \ i =1,2,3, \qquad 
\Gamma _4 = \begin{pmatrix}
0 & i \\
-i & 0
\end{pmatrix} ,
\end{equation} 
where $\{ \sigma _i \} $ are the Pauli matrices, we calculate
\begin{equation}
\begin{split} 
\det (\mathcal{I} _{ 12 })&=\det (\mathcal{I} _{ 34 })
=\mathcal{I}  + (K _1 ^2 + K _2 ^2 )\mathcal{H} ,\\
\det (\mathcal{I} _{ 13 })&=\det (\mathcal{I} _{ 24 })
=\frac{ \mathcal{I} }{16} + \left( K _1 ^2 + \frac{ K _2 ^2}{4} \right) \mathcal{H} ,\\
\det (\mathcal{I} _{ 14 })&=\det (\mathcal{I} _{ 23 })
=\frac{ \mathcal{I} }{16} + \left( K _2 ^2 + \frac{ K _1 ^2}{4} \right) \mathcal{H} ,
\end{split} 
\end{equation} 
where
\begin{align}
\mathcal{I} &=- \left( \frac{D  ^2 r ^2 + D (B _+ - B _- )r  - B _+ B _- }{(r ^2 - N ^2 )^6 } \right) , \quad 
\mathcal{H} =-\frac{1}{16 N ^2 P ^2  }\left( \frac{D ^2\mathrm{e} ^{ \mp  2 \varphi }  }{ (r \mp  N )^6 \Delta (r) } \right),
\end{align} 
and, with the sign choice referring to $F  =\pm * F $,
\begin{equation}
\begin{split} 
B _1  &=(M \pm  N P )^2 - N ^2 (P + 1-4 N ^2   )^2 , \\
B _2 & =(M \mp  N P )^2 - N ^2 (P -1+4 N ^2  )^2 , \\
D &= 2P  (M \mp  N  \pm  4 N ^3 ).
\end{split} 
\end{equation} 
Therefore we get the necessary conditions
\begin{equation}
\label{necond} 
D =0, \quad   B _1 B _2 =0.
\end{equation} 

Since $P \neq 0 $,  $D =0 $ is equivalent to
\begin{equation}
\label{ddcond} 
M =\pm  N (1 - 4 N ^2 ),
\end{equation} 
which is exactly the (anti) self-duality condition for the Weyl tensor, see  (\ref{msd}), (\ref{masd}), which we already knew to be a necessary condition for the existence of non-trivial Killing spinors. From now on we assume that  (\ref{ddcond}) is satisfied. More precisely, in the self-dual case we have $F =* F $, $M =\Msd $ and in the anti self-dual case $F =- * F $, $M =\Masd $.
Substituting (\ref{ddcond}) into the expressions for $B _1 $, $B _2 $  we see that
\begin{equation}
B _1 =B _2 =0 
\end{equation}
 automatically holds. 
Despite having considered a more general 2-form $F$, the conditions (\ref{necond}) are exactly the same  as those found in \cite{martelli:2013}. However,  because of the constraint  (\ref{detcond}), we find that  the system (\ref{odeq1}), (\ref{odeq2}) below admits no non-trivial solutions unless 
\begin{equation}
K _1 = K _2 =0 ,
\end{equation}
which is the case studied in \cite{martelli:2013}.  Since the treatment in \cite{martelli:2013} is mostly geared towards the non-self-dual case, for reference we reproduce their solutions in the self-dual case and provide some more details on the computation.

Write
\begin{equation}
\label{spinor} 
\epsilon =\begin{pmatrix}
\epsilon _1   \\
\epsilon _2  
\end{pmatrix} , \quad \epsilon _1  =
\begin{pmatrix}
\lambda _1  \\
\lambda _2 
\end{pmatrix}, \quad  \epsilon _2  =\begin{pmatrix}
 \mu _1 \\ \mu _2 
\end{pmatrix}.
\end{equation} 
For  $K _1 =K _2 =0 $,  (\ref{detcond}) gives 
\begin{align}
\label{mu1} 
\mu _1 &= i \left[
 \sqrt{ \frac{\Deltapm (r)  }{r^2  - N^2 }} \   \frac{ 1}{ r \pm \tfrac{N }{P } (1-4 N ^2 + P )}\right] ^{ \pm 1 }\lambda _1,\\
\label{mu2} 
\mu _2 &= i \left[
 \sqrt{ \frac{\Deltapm (r)  }{r^2  - N^2 }} \  \frac{ 1}{ r \mp  \tfrac{N }{P } (1-4 N ^2 - P )}\right] ^{ \pm 1 }\lambda _2.
\end{align} 
The sign choice corresponds to the self-dual or anti self-dual case and
\begin{equation}
\begin{split} 
\Delta _{ \pm }(r) &= (r\mp  N  ) ^2 \left( r-  x _\pm  \right) \left( r- y _\pm  \right),
\end{split} 
\end{equation} 
where $\ran$, $\rbn $ are given by (\ref{radtn}) and $\rab $, $\rbb $ by (\ref{rpmddd}).

In the self-dual case the radial component of the Killing spinor equation (\ref{kilspin})  is
\begin{align}
\label{odeq1} 
\partial _r \epsilon _1  &=- \frac{i}{2} \sqrt{ \frac{ r ^2 - N ^2 }{ \Deltatn (r) }}  \epsilon _2  ,\\
\label{odeq2} 
\partial _r \epsilon _2   &= +  \frac{i}{2} \sqrt{ \frac{ r ^2 - N ^2 }{ \Deltatn (r) }} \left[ I _2 + \left( \frac{ 2P}{(r+ N) ^2 } \right) \sigma _3  \right] \epsilon _1 ,
\end{align} 
and in the anti self-dual case
\begin{align}
\label{odeq1asd} 
\partial _r \epsilon _1  &=- \frac{i}{2} \sqrt{ \frac{ r ^2 - N ^2 }{ \Deltab (r) }} \left[ I _2 + \left( \frac{2P}{(r-N) ^2 } \right) \sigma _3  \right] \epsilon _2  ,\\
\label{odeq2asd} 
\partial _r \epsilon _2   &= +  \frac{i}{2} \sqrt{ \frac{ r ^2 - N ^2 }{ \Deltab (r) }} \  \epsilon _1  .
\end{align} 

In the self-dual case substituting (\ref{mu1}) in (\ref{odeq1}) and integrating the resulting ODE gives
\begin{align}
\label{lambda1} 
\lambda _1 &=   \sqrt{ r + N + \tfrac{N }{P } (1-4 N ^2 ) }\ \upsilon _1,\\
\label{lambda2} 
\lambda _2 &=   \sqrt{  r + N -\tfrac{N }{P }(1-4 N ^2 ) }\ \upsilon _2,
\end{align} 
with $\upsilon  _1 $, $ \upsilon  _2 $ at this point arbitrary functions of the angular coordinates. Substituting (\ref{lambda1}), (\ref{lambda2})  in (\ref{mu1}), (\ref{mu2})  gives
\begin{align}
\label{musol1} 
\mu _1 &
=i \sqrt{  \frac{\Deltatn (r) }{r ^2 - N ^2 }} \frac{1}{\sqrt{ r + N + \tfrac{N }{P } (1- 4 N ^2 )}} \ \upsilon _1 ,\\
\label{musol2} 
\mu _2 &
=i \sqrt{  \frac{\Deltatn (r) }{r ^2 - N ^2 }} \frac{1}{\sqrt{ r + N - \tfrac{N }{P } (1- 4 N ^2 )}}\  \upsilon _2.
\end{align} 
Equations (\ref{musol1}), (\ref{musol2}) are compatible with the solution of (\ref{odeq2}) with $\lambda_1 $, $\lambda _2 $ given by (\ref{lambda1}), (\ref{lambda2}) if and only if 
\begin{equation} 
\label{bpshalf} 
P ^2 = N^2  ( 4 N ^2 -1 ),
\end{equation} 
or
\begin{equation}
\label{bpsquarter} 
\left( P =\frac{4 N ^2 -1}{2} \text{ and } \lambda _2 =\mu _2 =0\right)  \quad \text{or}  \quad 
\left( P =\frac{1-4 N ^2 }{2} \text{ and } \lambda _1 =\mu _1 =0\right) .
\end{equation} 

Proceeding similarly, in the anti self-dual case  we find
\begin{align}
\mu  _1 &=   \sqrt{ r - N - \tfrac{N }{P } (1-4 N ^2 ) }\ \upsilon _1,\\
\mu  _2 &=   \sqrt{  r - N +\tfrac{N }{P }(1-4 N ^2 ) }\ \upsilon _2,\\
\lambda  _1 &
= - i \sqrt{  \frac{\Deltab (r) }{r ^2 - N ^2 }} \frac{1}{\sqrt{ r - N - \tfrac{N }{P } (1- 4 N ^2 )}} \ \upsilon _1 ,\\
\lambda  _2 &
= - i \sqrt{  \frac{\Deltab (r) }{r ^2 - N ^2 }} \frac{1}{\sqrt{ r - N +  \tfrac{N }{P } (1- 4 N ^2 )}}\  \upsilon _2.
\end{align} 
with the same compatibility conditions (\ref{bpshalf}), (\ref{bpsquarter}) found in the self-dual case.

The sign of $P$ can be changed by a coordinate redefinition, so without loss of generality we restrict to the two cases
\begin{align} 
\label{bpsh} 
P  &= -N  \sqrt{  4 N ^2 -1 },\\
\label{bpsq} 
 P &=\frac{1-4 N ^2 }{2} .
\end{align} 

Note that, for  anti self-dual bolt-type solutions, $N$ is given by (\ref{nspehcf}), so (\ref{bpsq}) becomes
\begin{equation}
\label{bpsquarterb} 
P = - \frac{1}{8} \, \frac{ (p-2 )^2 }{p-1 } ,
\end{equation} 
and (\ref{bpsh}) becomes
\begin{equation} 
\label{bpshalfb} 
P = - \frac{p}{8} \left(   \frac{ p-2  }{p-1 }\right)  .
\end{equation} 
Substituting (\ref{bpsquarterb}), (\ref{bpshalfb})  in (\ref{sdqcond}) we see that the quantisation condition arising in the case of anti self-dual bolt-type solutions is automatically satisfied provided that $P$ satisfies (\ref{bpsh}) or (\ref{bpsq}).

We now examine the angular part of the Killing spinor. Asymptotically,  to leading order in $r$,
\begin{equation}
g \simeq \frac{\mathrm{d} r ^2 }{r ^2 } + r ^2 g ^{ (3) }, \quad g ^{ (3) } = \eta _1 ^2 + \eta _2 ^2 + 4 N ^2 \eta _3 ^2 .
\end{equation} 
Let $ (e ^i )$ be the orthonormal coframe (\ref{on_cofr_axial}), $( e ^{ (3) } ) ^1 = \eta _1 $, $(e ^{ (3) })^2 =\eta _2 $, $(e ^{ (3) })^3 =2 N \eta _3 $. Denote by $\omega _{ ij }$, $\omega ^{ (3) } _{ ij }$ the Levi-Civita connection 1-forms associated to $g$ and $g ^{ (3) }$ with respect to  $(e _i )$ and $(e ^{ (3) }_i) $. We have
\begin{equation}
\begin{split}
\omega _{ 12 } &=\omega ^{ (3) }_{ 12 }=(2 N ^2  -1 ) \eta _3 ,\\
\omega _{ 13 } &=\omega ^{ (3) }_{ 13 }=N \eta _2 ,\\
\omega _{ 23 }&=\omega ^{ (3) }_{ 23 } =- N \eta _1 ,\\
\omega _{ i 4 }&= e _i , \ i =1,2,3.  
\end{split} 
\end{equation}
It follows
\begin{equation}
\begin{split}
\nabla _a \epsilon  &
= e _a (\epsilon ) + \frac{1}{4}  \omega _{ij } (e _a )\Gamma ^i \Gamma ^j  \epsilon 
= \frac{1}{r} \left[ X _a (\epsilon)  + \frac{1}{4}  \omega^{ (3) } _{ij} (X _a )\Gamma ^i \Gamma ^j \right]  \epsilon  + \frac{1}{2} \Gamma _a \Gamma _4 \epsilon ,
\end{split}
\end{equation}  
where $(X _i )$ are the left-invariant vector fields on $SU (2) $ dual to (\ref{etas}),
\begin{equation}
\begin{split}
X _1 &= \sin \psi\,  \partial _\theta + \frac{\cos \psi }{\sin \theta } \left( \cos \theta \, \partial _\psi - \partial _\phi \right) ,\\
X _2 &= \cos \psi\,  \partial _\theta - \frac{\sin  \psi }{\sin \theta } \left( \cos \theta \, \partial _\psi - \partial _\phi \right) ,\\
X _3 &= \partial _\psi.
\end{split}
\end{equation} 

Using the representation (\ref{gammarep}) for the gamma matrices and noticing that $ (\sigma _a )$ generate the Clifford algebra $\mathrm{Cl} (0,3 )$ we get
\begin{equation} 
\begin{split} 
\nabla \epsilon _1  &= \frac{1}{r} \left[ e ^{ (3) }_a (\epsilon _1 )  + \frac{1}{4}  \omega^{ (3) } _{ ij} (X _a )\sigma  _i \sigma  _j \right]  \epsilon_1   - \frac{i}{2} \sigma  _a  \epsilon _1 
=\frac{1}{r} \nabla_a  ^{ (3) } \epsilon _1 - \frac{i}{2}\sigma _a \epsilon _1  ,\\ 
\nabla \epsilon _2  &= \frac{1}{r} \left[ e ^{ (3) } _a (\epsilon _2 )  + \frac{1}{4}  \omega^{ (3) } _{ ij } (X _a )\sigma  _i \sigma  _j \right]  \epsilon_2 +  \frac{i}{2} \sigma  _a  \epsilon _2
=\frac{1}{r} \nabla_a  ^{ (3) } \epsilon _2  +  \frac{i}{2}\sigma _a \epsilon _2 .
\end{split}
\end{equation} 
We also define
\begin{equation}
A ^{ (3) } _a = \lim _{ r \rightarrow \infty }A (e ^{  (3) } _a ).
\end{equation} 
The components $F _{ ij } $ of $F =\mathrm{d} A $  with respect to the orthonormal frame $(e _i )$  are $O (r ^{ -2 })$ so to the leading order in $r$ they can be neglected. Thus asymptotically (\ref{kilspin})  becomes
\begin{align}
\label{asang1} 
(\nabla  ^{ (3) }_a  - i A ^{ (3) }_a )\epsilon  _1 + \frac{r}{2} \sigma _a (\epsilon _2 - i \epsilon _1 ) &=0,\\
\label{asang2} 
(\nabla  ^{ (3) }_a  - i A ^{ (3) }_a )\epsilon  _2 + \frac{i r}{2} \sigma _a (\epsilon _2 - i \epsilon _1 ) &=0.
\end{align} 

We now solve (\ref{asang1}), (\ref{asang2}) for $P$ given by (\ref{bpsq}) or (\ref{bpsh}). First consider the case
\begin{equation}
P = -N \sqrt{ 4 N ^2 -1 }
\end{equation} 
in the self-dual case. One has
\begin{equation}
\label{onehalfsol} 
\begin{split}
\epsilon  _1 &=\begin{pmatrix}
 \sqrt{ r + N +  \sqrt{ 4 N ^2 - 1}}\  \upsilon _1   \\
\sqrt{  r + N -\sqrt{4 N ^2 - 1}} \ \upsilon _2  
\end{pmatrix}
 ,\\
\epsilon _2  &= i\begin{pmatrix}
  \sqrt{ \frac{\Delta (r)  }{r ^2 - N ^2  }}  \frac{ 1 }{ \sqrt{ r + N + \sqrt{4 N ^2 - 1} }} \ \upsilon _1    \\
  \sqrt{ \frac{\Delta (r)  }{r ^2 - N ^2  }}  \frac{ 1 }{ \sqrt{ r + N - \sqrt{4 N ^2 - 1} }} \ \upsilon _2   
\end{pmatrix} .
\end{split}
\end{equation} 
Asymptotically, setting
\begin{equation}
\upsilon =\begin{pmatrix}
\upsilon _1  \\
\upsilon _2 
\end{pmatrix} ,
\end{equation} 
we get
\begin{equation}
\epsilon _2 - i \epsilon _1  \simeq -\frac{i}{\sqrt{ r }} (N I _2 +  \sqrt{4 N ^2 - 1} \,  \sigma _3 )\upsilon .
\end{equation} 
Both (\ref{asang1}) and (\ref{asang2}) thus reduce to the same equation
\begin{equation}
\label{asangeq} 
\left( \nabla ^{ (3) } _a - i A ^{ (3) } _a  - \frac{i}{2} N \sigma _a - \frac{i}{2} \sqrt{4 N ^2 - 1}\sigma _a \sigma _3  \right) \upsilon =0.
\end{equation} 
A similar computation shows that  the anti self-dual case also leads to equation  (\ref{asangeq}).

The general solution to (\ref{asangeq}) is \cite{martelli:2013,imamura:2012}
\begin{equation} 
\label{usol12} 
\upsilon =
\begin{pmatrix}
\cos \tfrac{\theta }{2}  \, \mathrm{e} ^{ \frac{i}{2} (\psi + \phi )}& - \sin \tfrac{\theta }{2} \, \mathrm{e} ^{ \frac{i}{2}  (\psi - \phi )}\\
 \gamma  \sin \tfrac{\theta }{2}\,  \mathrm{e} ^{- \frac{i}{2}  (\psi - \phi )} & \gamma \cos \tfrac{\theta }{2}\,   \mathrm{e} ^{ -\frac{i}{2} (\psi + \phi )}
\end{pmatrix} \upsilon _0 ,
\end{equation} 
with $\upsilon _0 $ a constant 2-spinor and
\begin{equation}
\gamma =i \left( 2 N + \sqrt{  4 N ^2 -1 }\right) .
\end{equation} 
One can check that the spinor (\ref{onehalfsol}), or the corresponding expression in the anti self-dual case, with $\upsilon$ given by (\ref{usol12}) solves all the components of the Killing spinor equation (\ref{kilspin}), thus giving a 1/2 BPS solution.

Take now
\begin{equation}
P =\frac{1-4 N ^2 }{2}.
\end{equation} 
Then in the self-dual case
\begin{equation}
\label{onequartttt} 
\begin{split}
\epsilon _1 &=\begin{pmatrix}
\sqrt{ r + 3N }\ \upsilon _1  \\
\sqrt{ r - N }\ \upsilon _2 
\end{pmatrix} ,\\
\epsilon _2 &= i\begin{pmatrix}
  \sqrt{ \frac{\Delta (r)  }{r ^2 - N ^2  }}  \frac{ 1 }{ \sqrt{ r + 3N  }} \ \upsilon _1   \\
  \sqrt{ \frac{\Delta (r)  }{r ^2 - N ^2  }}  \frac{ 1 }{ \sqrt{ r - N }} \ \upsilon  _2 \\
\end{pmatrix} .
\end{split} 
\end{equation} 
Asymptotically
\begin{equation}
\epsilon _2 - i \epsilon _1 \simeq \frac{i N }{\sqrt{ r }} \begin{pmatrix}
-3\upsilon _1  \\
 \upsilon _2 
\end{pmatrix} ,
\end{equation}
so both (\ref{asang1}) and (\ref{asang2}) reduce to
\begin{equation}
\label{asymkil} 
\left[ \nabla ^{ (3) } _a - i A ^{ (3) }_a \right] 
\begin{pmatrix}
\upsilon _1 \\
\upsilon _2 
\end{pmatrix} 
+ \frac{i N }{2} \sigma _a \begin{pmatrix}
 -3\upsilon _1  \\
  \upsilon _2 
\end{pmatrix}  =0.
\end{equation} 
Again, the anti self-dual case  leads to the same equation.

We have 
\begin{equation}
A ^{ (3) }_a =\left( \frac{P }{2 N }  \right)\delta _a ^3=\left(  \frac{1-4 N ^2}{4 N } \right) \delta _a ^3 ,
\end{equation} 
and, for $\upsilon =(\upsilon  _1 , \upsilon _2 )^T $,
\begin{equation}
\begin{split} 
\nabla _1 \upsilon  &=X _1 (\upsilon  ) - \frac{i N }{2} \sigma _1 \upsilon,\\
\nabla _2 \upsilon  &=X _2 (\upsilon  ) - \frac{i N }{2} \sigma _2 \upsilon,\\
\nabla _3 \upsilon  &=\frac{1}{2 N }X _3 (\upsilon  ) +  \frac{i (2N^2 -1 ) }{4N} \sigma _3 \upsilon.
\end{split} 
\end{equation} 
Therefore (\ref{asymkil}) is equivalent to the system
\begin{equation}
\label{oooonnnqqqq} 
\begin{split}
X _1 (\upsilon _1 ) &=X _2 (\upsilon _1 ) =X _3 (\upsilon_2   )=0,\\
X _1 (\upsilon _2 ) &=2i N  \upsilon _1 ,\\
X _2 (\upsilon _2 ) &=-2N  \upsilon _1 ,\\
X _3 (\upsilon _1 )&=i \upsilon _1  ,
\end{split}
\end{equation} 
which has general solution 
\begin{equation}
\label{usol14} 
\upsilon = \begin{pmatrix}
 0 \\
 \upsilon _0   
\end{pmatrix} ,
\end{equation} 
for $ \upsilon _0   $ a constant. One can check that the spinor (\ref{onequartttt}), or the corresponding expression in the anti self-dual case, with $\upsilon$ given by (\ref{usol14}) solves all the components of the Killing spinor equation (\ref{kilspin}), thus giving a 1/4 BPS solution.

To summarise, supersymmetry is preserved only if the Weyl tensor of the metric $g$ and the 2-form $F$  are both either self-dual or anti self-dual. In the self-dual case,  $F = P \beta _3 ^+ $, with $ \beta _3 ^+ $ given by (\ref{beta3plus}),  and $g$ is the  metric   of TN-AdS, see (\ref{tnmetric}), depending on one continuous  parameter  $N \in (0, \infty )$. Despite the  underlying manifold being topologically  trivial, Killing spinors  only arise if $P$ satisfies (\ref{bpsh}), leading to a 1/2 BPS solution, or (\ref{bpsq}), leading  to a 1/4 BPS solution.
In the anti self-dual case, $F =P \beta _3 ^- $, with $ \beta _3 ^- $ given by (\ref{beta3minus}),  and the  metric is of bolt-type, see (\ref{qeh}), depending on one discrete parameter $p\in \mathbb{Z}  $, $p \geq 3 $. The value of $P$ is again constrained by (\ref{bpsh}) and (\ref{bpsq}), leading to 1/2 and 1/4  BPS solutions as in the self-dual case. In this case the underlying manifold has a non-trivial topology and (\ref{bpsh}), (\ref{bpsq})  imply the quantisation conditions arising from the requirement that $F$ is the curvature of a $U (1) $ connection for a spin structure (for $p$ even) or $\mathrm{spin} ^{ \mathbb{C}  }$ structure (for $p$ odd).

\section{Conclusions}
In this paper we have determined a basis for the  space  of $SU (2) $-invariant  $L ^2 $ harmonic 2-forms on a 2-parameter family of bi-axial Bianchi IX Einstein metrics with negative cosmological constant $\Lambda $. In the case of NUT-type metrics,  this space is 3-dimensional, comprising entirely of self-dual forms. Since the underlying manifold is topologically trivial, these forms are necessarily exact. They are not, however, $L ^2 $-exact. In the case of bolt-type solutions, the space is 4-dimensional, with a 1-dimensional cohomologically non-trivial subspace and a 3-dimensional self-dual subspace.

These results can be viewed as a generalisation of those in \cite{franchetti2019harmonic}, which considers the Ricci-flat case $\Lambda =0 $. A negative cosmological constant results in a very different asymptotic behaviour of the metric, which is asymptotically hyperbolic rather than asymptotically locally flat. As a consequence, the space of $L ^2 $ harmonic 2-forms is infinite dimensional. Even the finite-dimensional subspace obtained by focusing on $SU (2) $-invariant forms is bigger than the  space of $L ^2 $ harmonic 2-forms obtained in \cite{franchetti2019harmonic} for $\Lambda = 0 $. The latter is 1-dimensional (respectively 2-dimensional) in the case of NUT type (bolt type) metrics. A basis for it can be obtained by taking the   $\Lambda \rightarrow 0 $   limit  of  the forms $ \beta _3  ^+ $ (respectively $ \beta _3 ^+  $ and $\beta _3 ^- $) found in this paper. The forms $\beta _1 ^+ $, $\beta _2 ^+ $ diverge in the limit.

A self-dual or anti self-dual  2-form $F$ on an Einstein manifold with cosmological constant provides a solution of the  bosonic sector of $4D$, $\mathcal{N} =2 $ minimal gauged  supergravity. It is therefore natural to ask whether any of the 2-forms that we have found gives raise to supersymmetric solutions. While we have shown that the question has a positive answer, all the solutions that we have found had been obtained before \cite{martelli:2013}. In fact, having started with  a more general form for $F$, which is only requested to be $SU (2) $-invariant rather than $SU (2) \times U (1) $-invariant, we have found that  Killing spinors only arise if $F$ is also $U (1) $-invariant, thus reducing the problem to the one studied in  \cite{martelli:2013}.

The lack of new solutions arises because of the mismatch between the $SU (2) $ invariance of the  2-form and the $SU (2) \times U (1) $ invariance of the metric. It would therefore be interesting to consider the case in which  both the 2-form and the metric   are  $SU (2) $-invariant only. 

\section*{Acknowledgements}
GF thanks the Simons Foundation for its support under the Simons Collaboration on Special Holonomy in Geometry, Analysis and Physics (grant 488631).

 \newpage 
 \bibliographystyle{amsplain}
\bibliography{biblio_paper}
\end{document}